\documentclass[10pt,twocolumn,twoside]{JCNtran}

\usepackage{graphics}
\usepackage[latin1]{inputenc}
\usepackage[T1]{fontenc}
\usepackage{algorithm} 
\usepackage{graphicx, amssymb}
\usepackage{latexsym,makeidx}
\usepackage{algorithmic}
\usepackage{tabularx}
\usepackage{textcomp,booktabs}
\usepackage{multirow}
\usepackage{array}
\usepackage{amsmath}
\usepackage{bm}

\def\BibTeX{{\rm B\kern-.05em{\sc i\kern-.025em b}\kern-.08em
    T\kern-.1667em\lower.7ex\hbox{E}\kern-.125emX}}

\begin{document}
\bibliographystyle{jcn}

\title{Energy-aware Traffic Engineering in Hybrid SDN/IP Backbone Networks}

\author{Yunkai Wei, Xiaoning Zhang*, Lei Xie and Supeng Leng
\thanks{Manuscript received October 23, 2015; approved for publication by Grace Kim, April 23, 2016.}
\thanks{This work has been supported by National Key Technology Research and Development Program of China under Grant No.2015BAH08F01, the National Natural Science Foundation of China (61201129), Information Technology Research Projects of Ministry of Transport of China (2014 364X14 040), and the Fundamental Research Funds for the Central Universities (ZYGX2013J009).}
\thanks{The Authors are with the School of Communication and Information Engineering, University of Electronic Science and Technology of China, Chengdu, China. Postcode: 611731. Xiaoning Zhang is the corresponding Author. Email: \{ykwei, xnzhang, lxie, spleng\}@uestc.edu.cn.}} 

\markboth{JOURNAL OF
COMMUNICATIONS AND NETWORKS, VOL. x, NO. x, xxx
2016}{Wei \lowercase{\textit{et al}}.: Energy-aware Traffic Engineering in Hybrid SDN/IP Backbone Networks} \maketitle

\begin{abstract}
Software Defined Networking (SDN) can effectively improve the performance of traffic engineering and has promising application foreground in backbone networks. Therefore, new energy saving schemes must take SDN into account, which is extremely important considering the rapidly increasing energy consumption from Telecom and ISP networks. At the same time, the introduction of SDN in a current network must be incremental in most cases, for both technical and economic reasons. During this period, operators have to manage hybrid networks, where SDN and traditional protocols coexist. In this paper, we study the energy efficient traffic engineering problem in hybrid SDN/IP networks. We first formulate the mathematic optimization model considering SDN/IP hybrid routing mode. As the problem is NP-hard, we propose the fast heuristic algorithm named HEATE (Hybrid Energy-Aware Traffic Engineering). In our proposed HEATE algorithm, the IP routers perform the shortest path routing using the distribute OSPF link weight optimization. The SDNs perform the multi-path routing with traffic flow splitting by the global SDN controller. The HEATE algorithm finds the optimal setting of OSPF link weight and splitting ratio of SDNs. Thus traffic flow is aggregated onto partial links and the underutilized links can be turned off to save energy. By computer simulation results, we show that our algorithm has a significant improvement in energy efficiency in hybrid SDN/IP networks.
\end{abstract}


\section{\uppercase{Introduction}}
\label{sec:introd}
\PARstart{S}{oftware} Defined Network (SDN) is a new networking paradigm [1-3] with the control plane and the data plane separated, which allows operators to easily deploy network applications through a central controller in the control plane and to distribute the fine-grained policies into the switch flow table in the data plane through standard interfaces. In the SDN, a logically centralized controller that has a global network state is responsible for path selection and it communicates with the network-wide distributed forwarding elements by OpenFlow protocol. Google has reported that it is using a SDN to interconnect its world-wide data centers due to the ease, efficiency in achieving traffic engineering objectives [4]. It expects that the SDN architecture to realize better network capacity utilization and reduce time delay.
\par 
At present, fulfilling full SDN deployment is impossible for network operators in a short term as it requires a lot of modifications to current existing networking architecture and the replacement of "old" network equipments cause high cost. On the other hand, SDN comes with its own set of challenges and limitations, ranging from deployment obstacles to concerns on logic centralization guarantees, e.g., in terms of resilience, robustness and scalability. Thus the introduction of SDN in an existing network must be incremental in most cases. Hybrid SDN model will be widely used in the future. As shown in Fig.1, in a hybrid SDN/IP network, the SDN-enabled switches and traditional IP routers co-exist, which are distinguished by whether supporting SDN protocols. The detailed description of the hybrid model is provided in Subsection III.B.
\par 
At the same time, energy consumption from the Information and Communication Technologies (ICT) sector, and in particular from Telecom and ISP networks, is increasing fast. To mitigate this problem, new schemes must adapt to such new environment of hybrid SDN/IP networks.
\par 
In this paper, we study the problem of energy-aware traffic engineering in hybrid SDN/IP networks. Our objective is to minimize network energy consumption. In our considered hybrid network, we assume only one controller controls all SDN-enabled switches in the network, and the rest of the network nodes are IP routers running hop-by-hop routing using a standard routing protocol. In IP network, Open Shortest Path First (OSPF) is the most commonly used intra-domain internet routing protocol. The network operator assigns a weight to each link, and traffic flows are routed along the shortest path which are computed using these weights. IP network energy can be minimized by finding an optimal set of link weights [5]. In such a hybrid SDN/IP network, we propose energy-aware traffic engineering algorithm that the set of OSPF link weights and traffic flow routing controlled by the SDN controller can be jointly optimized to achieve energy efficiency. We hope that the proposed algorithm can adaptively and dynamically manage traffic in a network to accommodate traffic varies and minimize network energy consumption. In general, the main contributions of our work can be summarized as follows:
\par 
- We first present a linear programming mathematic model to minimize the SDN/IP hybrid network energy. In this model, IP routers perform the shortest path routing by the set of link weight from OSPF protocol; SDN-enabled switches perform multi-path routing with traffic split by the central SDN controller.
\par 
- We also propose the fast heuristic algorithm named HEATE (Hybrid Energy-Aware Traffic Engineering). The HEATE algorithm mainly consists of two parts: the first part is the link weight optimization for IP routers based on Neighboring Region Search and the second one is traffic flow splitting at the SDN-enabled switches by the global controller. The algorithm aggregates traffic flow onto partial links and turns off underutilized links for energy saving.
\par 
- We conduct extensive simulations to evaluate our proposed algorithm using real network topologies and traffic traces. We find that even when only a small number of SDN-enabled switches are deployed in the traditional IP network, a significant improvement in energy efficiency can be gained.
\par 
The rest of the paper is organized as follows. Section \ref{Section:related_work} briefly describes the related work. The hybrid SDN/IP network architecture is presented in Section III. In Section IV, we formulate a mathematical optimization model to minimize energy consumption in the hybrid network. The fast HEATE heuristic algorithm is presented in Section V. The simulation results and the performance analysis are depicted in Section VI. Finally, conclusions are made in Section VII. 

\vspace{10pt}
\section{\uppercase{Related Work}}
\label{Section:related_work}
There have been a number of works focused on saving energy in traditional networks. The energy saving problem of OSPF protocol is studied in [6], where a novel network-level strategy is proposed to save energy during low traffic periods. The solution in [6] is that only a subset of router Shortest Path Trees (SPTs) are used to select the routing paths. The works in [7-9] focused on saving energy of networks in other scenarios. By selectively powering down individual cables of large bundled links, [9] developed and evaluated techniques that save energy in core networks. Several easy-to-implement heuristics are also developed in this work. In addition, Literature [7] and [8] studied energy saving in mobile radio networks and cellular access networks, respectively, and developed their own heuristic algorithms as well.
\par 
Recently, some works studied the problem of saving energy in software-defined networks. Lots of effort has gone into estimating the energy consumption at the different parts of a network, including core, transport and access [10-14], data centers [15-16], and end-user devices [17]. Based on OpenFlow, literature [18] focused on saving energy without explicit constraints, where they dynamically detected switches with minimal trafic and powered them off by consolidating the flows to other switches from them. Literature [19] presented an OpenFlow controller that created a loop-free layer-2 topology and reduced the network energy consumption by switching off inactive interfaces.
\par 
To the best of our knowledge, this is the first paper developing energy-aware traffic engineering techniques for hybrid SDN/IP networks. Most of works in this area so far are focused on devising pure SDN, and ignore hybrid SDN architecture and corresponding application. A small number of works [20-22] studied the hybrid SDN network. In [20], four hybrid SDN models (i.e., Topology-Based, Service-Based, Class-Based and Integrated) are defined and explored, and corresponding use cases (including transition and long-term design use cases) are described. The authors of [20] also provided a comparative analysis of the presented hybrid SDN models and showed that these models can mitigate the respective limitations of traditional and SDN ways. The authors of [21] considered traffic engineering in the case where a SDN controller controls only a few SDN forwarding elements in the network. The paper formulated the SDN controllers' optimization problem for traffic engineering with partial deployment and developed fast Fully Polynomial Time Approximation Schemes (FP-TAS) for improving network capacity utilization. In [22], the authors developed techniques to adapt traffic flows to network varies, anomaly-free update routing policies and incrementally deploy SDN in traditional networks. These techniques enable traffic engineering and policy routing at a fine-grained level.
\par 
The above works of hybrid SDN networks aim to maximize the network capacity utilization or reduce packet loss rate during flow table update. Our paper first researches on improving energy efficiency for the hybrid SDN/IP network. To gain better energy-saving effect in SDN/IP hybrid networks, we jointly optimize the set of link weights among IP routers and traffic flow splitting ratio of SDN-enabled switches.

\vspace{10pt}
\section{\uppercase{The Hybrid SDN/IP Network Architecture}}
\label{Section:system_model}
This section presents the SDN/IP hybrid network architecture which is the co-existence of traditional environment with SDN-enabled switches. First, we briefly describe SDN architecture. Then the SDN/IP hybrid network model and routing mechanism are demonstrated respectively.

\subsection{SDN architecture}
SDN architecture includes three layers: data-plane layer, control-plane layer and application layer. North-bound APIs are used to communicate the control-plane layer with application layer and enable application layer provide different patterns of applications, such as routing calculation, traffic engineering, topology measurement, etc. The controller in control-plane layer collects the network information and manages flows going through SDN-enabled switches in the data-plane layer via South-bound APIs.
\par 
The details of flow management are explained as follows. When a flow arrives at a SDN-enabled switch and cannot match any flow entry in the flow table, the following actions will be carried out: (1) the first packet of the flow is sent by the switch to the controller, (2) the controller selects the forwarding path from the candidate paths for the flow, (3) the controller sends the calculated flow entries to the flow table at each switch along the appropriate path, (4) all subsequent packets in the flow that can match the flow entries in the flow table, are forwarded in the data plane along the path and do not need to be sent to the controller again. From the above actions, the routing flow tables at the SDN-enabled switches are computed by the controller. Through the three-layer SDN architecture, central and fine-grained control for the network management over the traffic flows can be realized in a global network perspective.

\subsection{The SDN/IP hybrid network model}
Similar to the previous works [21, 22], in this paper, we define the hybrid model on basis of Route Flow architecture [23]. The hybrid SDN model combines SDN-enabled devices with traditional existing networks. That is, the data forwarding plane consists of traditional IP routers and SDN-enabled switches. We distinguish between traditional IP routers and SDN-enabled switches according to whether support SDN protocols. The SDN-enabled switches are controlled by a SDN controller, which means the forwarding flow table of the SDN-enabled switches is computed by the controller. The IP routers use standard hop-by-hop routing protocol like OSPF to forward data packets. To focus on the energy aware traffic engineering problem, we assume the SDN-enabled switches are already deployed, randomly and uniformly distributed. An example of hybrid SDN/IP network is depicted in Fig.1. Solid lines represent data links which are used to forward data traffic in the data forwarding plane, while dashed lines represent channels used to send control traffic (traffic to or from the controller). We assume that all the data links in the network are bidirectional and all link OSPF weights are set to 1. Nodes 3, 8, 10 are SDN-enabled switches which are controlled by a SDN controller.

\begin{figure}[t]
\begin{center}
\includegraphics[width=0.48\textwidth]{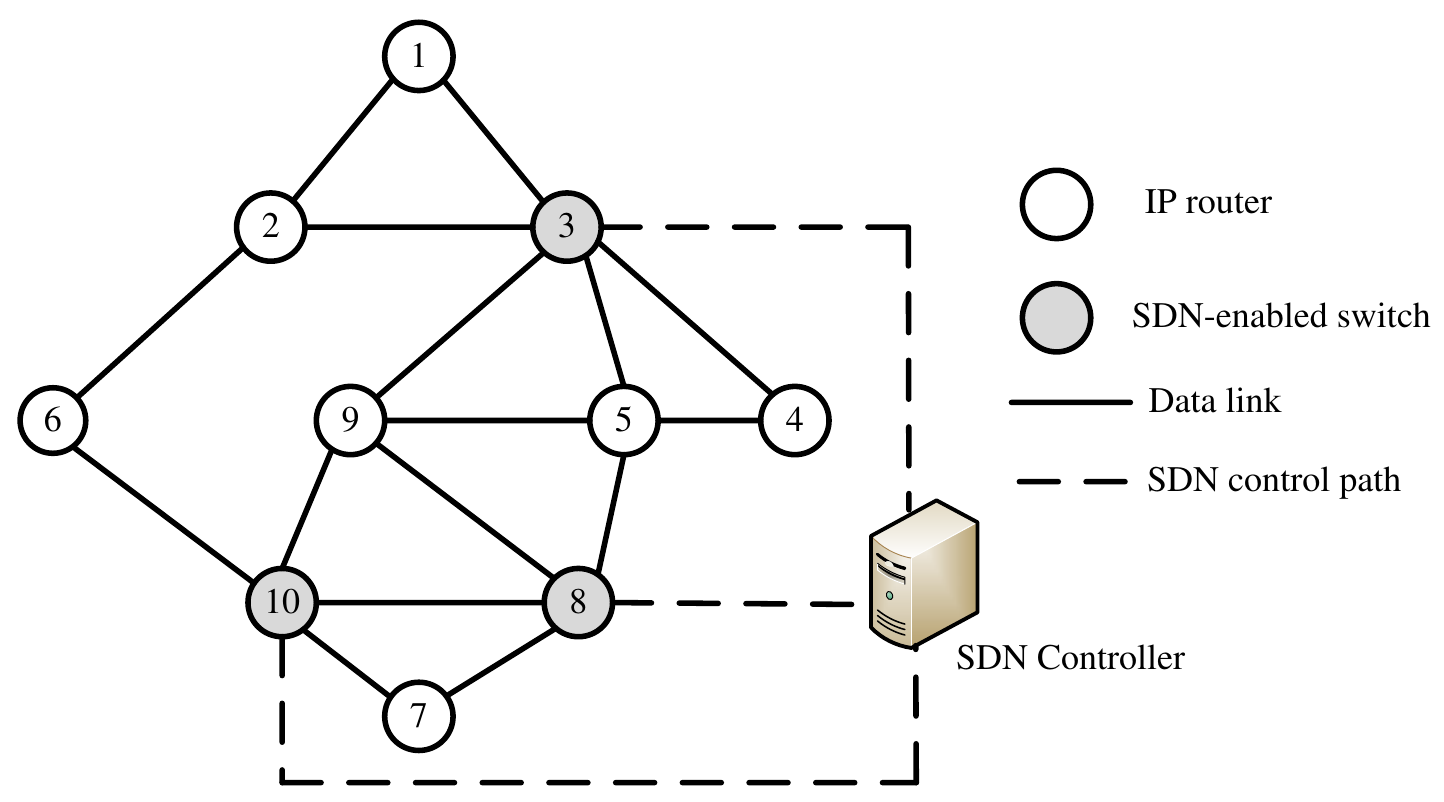}
\caption{An Example of IP/SDN Hybrid Network} 
\label{Figure:hybrid_example}
\end{center}
\end{figure}

\subsection{The SDN/IP hybrid routing mechanism}
From the perspective of traditional IP routers, they regard the SDN-enabled switches as common IP routers. So when computing traffic forwarding paths of the traditional IP routers, the existence of SDN-enabled switches are externally transparent to other IP routers.  The IP routers run OSPF protocol, establish adjacency relation between them by sending a Hello message, exchange network topology information such as link weights by utilizing Link State Advertisement (LSA) message, and update Link-State Database (LSD) which keeps the whole network topology information. Finally all the IP routers maintain the same LSD, compute the shortest path trees, and establish routing table and Forwarding Information Base (FIB) according to the LSD. Meanwhile, SDN-enabled switches send the network topology information to the SDN controller. The SDN controller knows the current OSPF weights as well as the amount of traffic flows on each link. It computes the traffic routing based on the information. The SDN-enabled switch may has multiple next-hops for a destination. In Fig.2, we give an example of the routing path from all other nodes to node 5 for the network model of Fig.1. The solid lines represent the shortest path tree based on OSPF protocol to switch 5 and the dotted lines represent the feasible path from SDN-enabled switches. For an IP router, there is only one feasible path to node 5 if ECMP (Equal Cost Multiple Paths) does not exist to the destination. For a SDN-enabled switch, there can be multiple paths that are associated with the outgoing links of the SDN-enabled switch.

\begin{figure}[t]
\begin{center}
\includegraphics[width=0.48\textwidth]{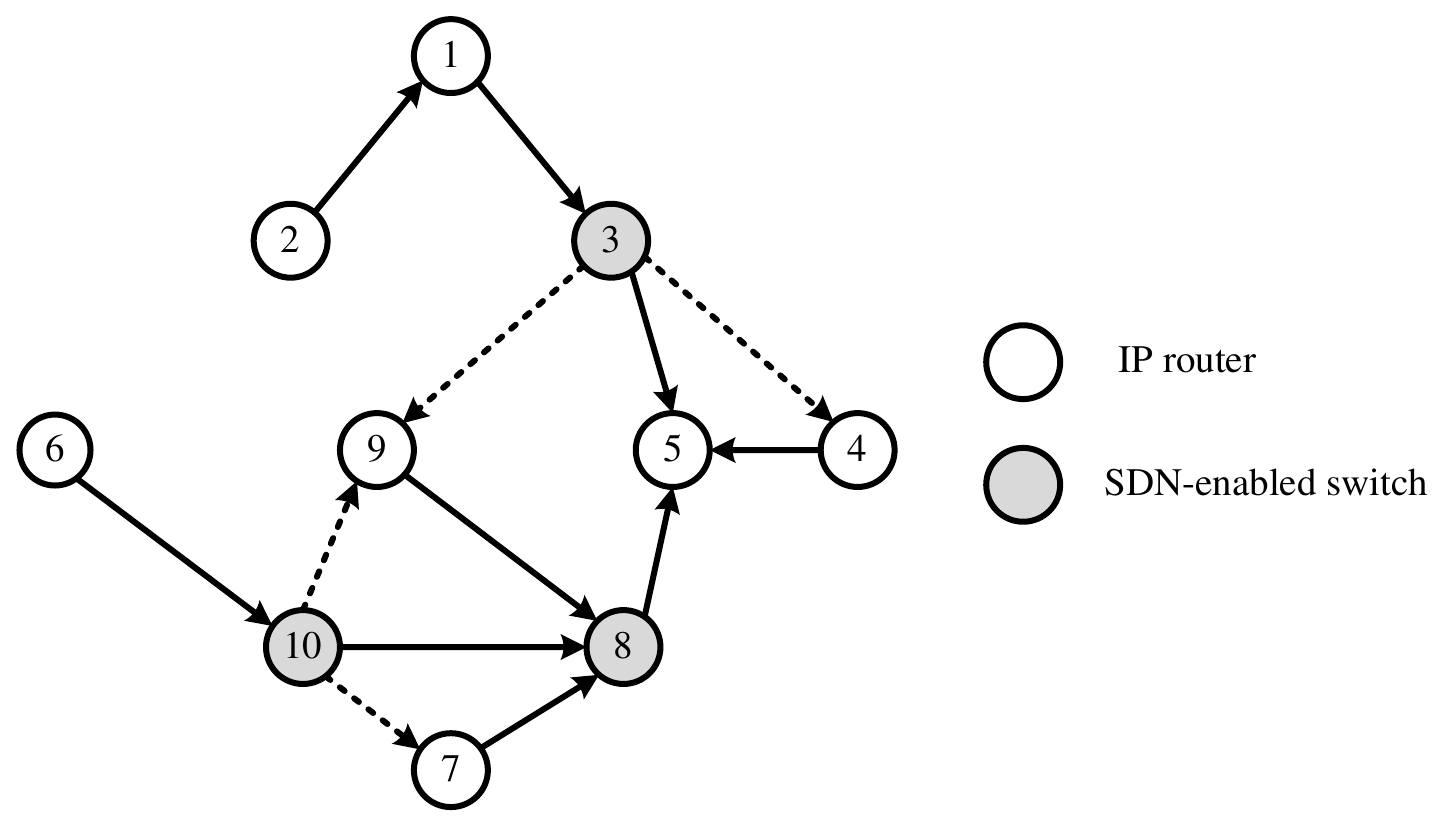}
\caption{The Shortest Path to Node 5} 
\label{Figure:shortest_path}
\end{center}
\end{figure}

\vspace{10pt}
\section{\uppercase{The Energy-Aware Traffic Engineering Optimization Problem}}
A key factor of energy-aware traffic engineering problem is the network energy consumption model. There are two popular models to determine network energy consumption. One is powering down model, and the other is speed scaling model [24]. The former reduces energy consumption by turning off unnecessary network elements, such as line cards and routers, and the latter achieves energy saving by setting the processing speed of the network elements according to the traffic load.  We focus on the powering down model in this paper. In this model, each network element has two opposite state: the active state at the full rates and the sleep state at the zero rates. The network element consumes corresponding energy in the active state and does not consume energy in the sleep state.  We assume work elements consume the same energy when they operate in the active state. Thus we can achieve energy efficiency by aggregating traffic flows onto a subset of links and turning off idle links which no traffic flow traverses through.
\par 
\begin{table}
\caption{Notations used in the problem}
\begin{tabularx}{8.5cm}{>{\hsize=0.15\hsize}X >{\hsize=0.85\hsize}X}
\hline
\toprule
Notation & ~Description\\
\hline
\midrule

$G(N,L)$ & The hybrid network {\bf\em G}, where {\bf\em N} is the set of network nodes and {\bf\em L} is the set of directed physical links.\\

$v,u,t$ & The node in the hybrid network, and {\em v,u,t} $\in$ {\bf\em N}.\\

$w_l$ & The weight of link $l$.\\

$x_l$ & The total traffic flow of link $l$.\\

$c_l$ & The capacity of link $l$.\\

$\mu_l$ & The utilization of link $l$.\\

$M$ & A large number.\\

$\beta$ & The maximum link utilization.\\

$s_l$ & The starting node of link $l$.\\

$d_l$ & The terminating node of link $l$.\\

$h_{vt}$ & The traffic demand from node {\em v} to some other node {\em t}.\\

$I_{ut}$ & The traffic flows that are injected at a SDN-enabled switch $u$ to some destination {\em t}.\\

$r_{vt}$ & The sum of weight for the shortest path from node $v$ to node $t$.\\

$x_{lt}$ & The flow to node $t$ on link $l$.\\

$y_{vt}$ & The common value of non-zero flow from node $v$ to node $t$ assigned to links outgoing from $v$ and belonging to the shortest-paths from $v$ to $t$.\\

$u_{lt}$ & Binary variable. It equals to 1 if and only if link $l$ is on a shortest-path to node $t$.\\

$p(l)$ & Binary variable. It equals to 1 if link $l$ is selected to be an active link; otherwise, it equals to 0.\\

$C$ & The set of SDN switches in the network, and $C \subseteq N$.\\

$D$ & The set of traditional IP routers in the network, and $D\subseteq N$.\\
\bottomrule
\hline
\end{tabularx}
\label{Table:notation}
\end{table}

For clear presentation, we summarize the notations that will be used in the problem in Table \ref{Table:notation}. The energy-aware traffic engineering problem in hybrid SDN/IP networks consists of two parts: the OSPF link weight optimization at IP routers and traffic flow splitting at the SDN-enabled switches managed by the SDN controller. In the following, we first give the energy-aware traffic engineering optimization problem.
\par 
{\em {Minimize:}}
\begin{equation}
\sum_{l}p(l)
\end{equation}

{\em {Subject to:}}
\begin{equation}
\sum_{l:d_l=t}x_{lt}=\sum_{s\neq t}h_{st}~~~~~~\forall t\in N
\end{equation}
\begin{equation}
\sum_{l:s_l=v}x_{lt}-\sum_{l:d_l=v}x_{lt}=h_{vt}~~~~~\forall t\in N, v\in N, v\neq t
\end{equation}
\begin{equation}
x_l=\sum_{t}x_{lt}~~~~~~\forall l\in L
\end{equation}
\begin{equation}
x_l/c_l\leq\beta p(l)~~~~~~\forall l\in L
\end{equation}
\begin{equation}
0\leq y_{s_lt}-x_{lt}\leq (1-u_{lt})\sum_{v}h_{vt}~~~~\forall t\in N,l\in L,s_l\in D
\end{equation}
\begin{equation}
x_{lt}\leq u_{lt}\sum_{v}h_{vt}~~~~~~\forall t\in N,l\in L,s_l\in D
\end{equation}
\begin{equation}
r_{d_lt}+w_l-r_{s_lt}\leq (1-u_{lt})M~~~~~~\forall t\in N,l\in L,s_l\in D
\end{equation}
\begin{equation}
1-u_{lt}\leq r_{d_lt}+w_l-r_{s_lt}~~~~~~\forall t\in N,l\in L,s_l\in D
\end{equation}
\begin{equation}
\sum _{s_l=v}u_{lt}=1~~~~~~\forall t\in N,v\in D,v\neq t
\end{equation}
\begin{equation}
w_l\geq 1~~~~~~~~~\forall t\in N,l\in L
\end{equation}

\par 
The objective function is to minimize the total energy consumption in the network. Since we assume that every link consumes the same energy, the objective is converted to minimizing total number of active links in the hybrid network.  Constraint (2) and Constraint (3) are flow conservation constraints that ensure the traffic flows are routed from source node to destination node. Constraint (4) ensures the link load equals all traffic flows carried on it. Constraint (5) states physical capacity constraints that guarantee the link load does not exceed the threshold of link capacity. Constraint (6-10) guarantee traditional IP router has a unique next hop which is on the shortest path to destination and SDN switches may have multiple next hops. Constraint (6) assures that if link $l$ belongs to one of the shortest-paths from traditional IP router $s_l$ to node $t$, then its flow to node $t$ is equal to $y_{s_lt}$-a common value to all links outgoing from IP router $s_l$ and belonging to the shortest-paths to destination $t$. Constraint (7) assures the zero flow to $t~(x_{lt}=0)$ in the case when link $l$ is not on the shortest-path to $t$. Constraint (8) assures that if $u_{lt}$, then link $l$ is on the shortest-path to $t$. Constraint (9) assumes that if $u_{lt}=0$, then link $l$ is not on the shortest-path to $t$. Constraint (10) forces the shortest-path to be unique provided such a solution is feasible. Constraint (6-10) only guarantee the next hop of the traditional IP routers are on the shortest path to destination, and SDN-enabled switches may have multiple next hops which may be not on the shortest path. Constraint (11) restricts the range of link weights.
\par 
When the OSPF link weight is fixed, the problem boils down to a multi-commodity linear programming problem, which can be solved in polynomial time [25]. However, the OSPF link weights are undetermined, which adds up the complexity of the problem. It has been proved that optimizing of link weight setting is a NP-hard problem [26]. Our proposed model can be reduced to the OSPF link weight optimization problem if traffic splitting ratio on all SDNs is determined in advance. That is to say, comparing the constraints of the mathematic problem Eq.(1-11) with that of the linear program Eq.(1-9) in literature [26], we can find the latter is a subset of the former. Since the latter has already been proved to be a NP-hard problem in [26], the mathematic problem Eq.(1-11) in this paper is also a NP-hard problem.

\vspace{10pt}
\section{\uppercase{The HEATE Algorithm}}
\label{Section:HEATE}
Since the problem to solve in our work is NP-hard, we propose the fast heuristic HEATE (Hybrid Energy-Aware Traffic Engineering) algorithm that jointly optimizes link weights of IP routers and traffic splitting ratio at SDN-enabled switches to improve energy efficiency.
\par 
\begin{algorithm}[tbp]
\caption{HEATE Algorithm}
\label{Algorithm:HEATEAlg}
\begin{algorithmic}[1]
\REQUIRE $G(N,L)$, $h_{vt}$, $\beta$, initial link weight set $W$.
\ENSURE The set of active links {$S$}.
\STATE Initialize $S \gets L$;
\STATE Run \emph{Neighboring~Region~Search} function to update link weight $w_l$, where $l \in S$;
\FOR {all $t \in N$}
    \STATE Generate the shortest path tree from any other node to $t$ based on $w_l$, find the first SDN-enabled switch $u$ in the shortest path and compute the injected flow at SDN-enabled switch $I_{ut}$;
    \STATE Run \emph{Flow~Alloction} function at the SDN-enabled switch for $I_{ut}$;
\ENDFOR
\STATE Try to delete the minimum utilization link $l$ from the network. If successful, $S=S-l$ and go back to line 2; else,  \textbf {the algorithm ends and return $S$}.
\end{algorithmic}
\end{algorithm}

\subsection{The algorithm description}
The detail of HEATE algorithm is shown in Alg.\ref{Algorithm:HEATEAlg}. Since the OSPF link weight and traffic splitting ratio are interdependent, we only fix one parameter to calculate the other parameter in the algorithm. The core of the algorithm is \emph{Neighboring Region Search} function (line 2) and \emph{Flow Allocation} function (line 5). The input of the algorithm includes the initial OSPF link weight set $W$ (in our study, we assume the initial link weight equals 1). We first optimize link weight for energy efficiency using \emph{Neighboring Region Search} function (line 2). The detail of \emph{Neighboring Region Search} function is shown in Alg.2. Thus the shortest path tree from other nodes to a special node is generated based on new link weight set $W$. The injected flow to the first SDN-enabled switch on the path can be calculated. As SDN-enabled switch can split traffic to the destination $t$, the algorithm calls \emph{Flow Allocation} function to compute the traffic splitting ratio on outgoing links from the switch (line 3-6). The detail of \emph{Flow Allocation} function is shown in Alg.3. After determining the traffic flow on IP routers and SDN-enabled switches, we can obtain the utilization of all links. We try to delete the minimum utilization link from the network to save energy. If successful, which means the utilization of residual links is less than the utilization threshold $\beta$, we delete the link from the network and loop back to line 2; else the algorithm ends and the final result is returned.

\begin{algorithm}[tbp]
\caption{Neighboring Region Search Function}
\label{Algorithm:Neighbor_Region_Search}
\begin{algorithmic}[1]
\REQUIRE The initial link weight set $W$=\{$w_l$\}, The maximum link utilization $\beta$.
\ENSURE The new weight $w_l$ of link $l$.

\FOR {$i$:=1 to $Iteration$}
\STATE Compute the shortest path routing in the network based on the link weight set $W$=\{$w_l$\};
\STATE Compute the link utilization $u_l \gets x_l/c_l$;
\IF {$u_l \leq 0.3 \times \beta$ }
      \STATE $w_l \gets w_l+1/i$;
     \ELSE
       \IF {$ u_l \geq \beta $}
       \STATE  $w_l \gets w_l+1/i$;
       \ENDIF
   \ENDIF
\ENDFOR
\STATE \textbf{return} $w_l$.
\end{algorithmic}
\end{algorithm}

As shown in Alg.2, \emph{Neighboring Region Search} function is to determine the searching area and adjustment method when updating link weight. According to the relationship between link weights and traffic flows, the function adjusts the link weights based on link utilization. At each iteration of link weight adjustment, the function distinguishes network links based on the link utilization into three categories: congested links (the link utilization is bigger than $\beta$), sleeping links (the link utilization is less than $0.3\times \beta$), middle-utilization links (the other links). For congested links, the function increases the weights to reduce the probability of being demand forwarding paths. It reduces link utilization and avoids network congestion. For sleeping links, the function also increases the weights to reduce the probability of being forwarding paths so that the link utilization can move toward zero direction for energy saving. For each iteration, the method of link weight adjustment is based on Harmonic Series [27] to improve searching efficiency and the number of iterations in the function is set to 10,000.

\begin{algorithm}[tbp]
\caption{Flow Allocation Function}
\label{alg:groupTest}
\begin{algorithmic}[1]
\REQUIRE The shortest path from other nodes to node $t$, the traffic flow $x_l$ of all links of the path, the link capacity $c_l$ of all links of the path, the first SDN-enabled switch $u$ on the shortest path, the injected traffic flow $I_{ut}$, the link utilization threshold $\beta$.
\ENSURE The flow $I_{ut}$ allocation among the multiple paths.
\STATE For the first SDN-enabled switch $u$, if there are $k$ outgoing links from $u$, calculate the $k$-th shortest path from node $u$ to node $t$ and store them to set $P=\{p_{ut}\}$;
\FOR {all $p_{ut} \in P$}
   \FOR {all $l \in p_{ut}$}
      \STATE $cap_l=c_l \times \beta-x_l$;
   \ENDFOR
   \STATE $cap_p=arg min_{l\in P}(cap_l)$;
\ENDFOR
\STATE Sort all paths of $P$ in the ascending order of available path capacity $cap_p$;
\STATE Initialize: $i \gets 1$;
\WHILE {$I_{ut}>0$}
   \STATE Route the flow volume $cap_{P[i]}$ on the path $P[i]$;
     \FOR {$j=i+1$ to $k$}
       \IF {$P[j]$ and $P[i]$ is jointed}
         \STATE $cap_{P[j]}=cap_{P[j]}-cap_{P[i]}$;
       \ENDIF
     \ENDFOR
   \STATE $I_{ut} \gets I_{ut}-cap_{P[i]}$;
   \STATE $i \gets i+1$;
\ENDWHILE
\STATE \textbf{return} the flow allocation result for $I_{ut}$.
\end{algorithmic}
\end{algorithm}
\par 
As shown in Alg.3, Flow Allocation function performs traffic splitting among multiple next hops in the first SDN-enabled switch on the shortest path to node . To save network energy, the aim of the function is to move traffic flow from low-utilization links to high-utilization links. We first compute the shortest paths from the SDN-enabled switch   to the destination node , and then compute the available capacity of these paths (Line 1-7). We prefer to allocate traffic flow on high-utilization path (i.e., the path with minimal available capacity) to achieve flow aggregation. The flow allocation procedure continues until the flow volume  is completely allocated among multiple paths. Considering the shortest paths are not guaranteed to be disjointed, we use line 12-16 to calculate the real residual capacities of the jointed paths.
\par 
Obviously, the algorithm complexity of HEATE is $O(n^2)$, where $n$ is the number of the nodes in the set of network nodes $N$. Considering that the flow in backbone networks will not change dramatically, HEATE will not be called very frequently. Therefore, the energy consumption of running this algorithm can be ignored comparing to the saved energy by HEATE.

\subsection{A simple example of the algorithm}
We use Fig.3 as a simple example to illustrate our algorithm. The graph is a simple network topology. It has three nodes and node A is a SDN-enabled switch, and node B and C are both IP routers (see Fig.3(a)). We assume the link capacity is set to 10 units, and $\beta$ is set to 0.8. There are totally three traffic flows in the network. The traffic flows are given in the form (source, destination, bandwidth units): $T_1$=(A,B,1), $T_2$=(A,C,3) and $T_3$=(C,B,3). The initial link weight is set to 1. Fig.3(b)-(d) shows the procedure of \emph{Neighboring Region Search} Function in the algorithm. Fig.3(b) shows the link load based on initial OSPF link weight. Since the initial link weight is set to 1, the path of $T_1$ is A-B, the path of $T_2$ is A-C, and the path of $T_3$ is C-B. So the traffic load of link (A,B), (A,C) and (C,B) is 1 unit, 3 units and 3 units, respectively. Fig.3(c) shows the link load based on the first-iteration adjusted link weight. Since the load utilization of link (A,B) equals 0.1 and is less than the sleeping threshold ($0.3 \times \beta=0.24$), accordingly the weight of link (A,B) is changed to 2 (see line 5 of \emph{Neighboring Region Search} Function) and other link weight is unchanged. Then there exist two equal-cost paths from node A to node B: A-B and A-C-B. The traffic flow $T_1$ is equally spitted between two paths. Therefore, the traffic load of (A,B), (A,C) and (C,B) is 0.5 unit, 3.5 units and 3.5 units respectively. Fig.3(d) shows the link load based on the second-iteration adjusted link weight. Since the load utilization of link equals 0.05 and is less than the sleeping threshold ($0.3 \times \beta=0.24$), accordingly the weight of link (A,B) is changed to 2.5 (see line 5 of \emph{Neighboring Region Search} Function) and other link weight is unchanged. Then the shortest path from node A to node B is A-C-B. So the traffic load of link (A,B), (A,C) and (C,B) is 0 unit, 4 units and 4 units, respectively. 
\par 
We note that Neighboring Region Search Function aggregates traffic flow onto partial links by two iterations of link weight adjustment in this sample. For large-scale network, more iteration times are needed to have effect. Fig.3(e) shows the shortest path from other nodes to node B. The solid lines are the shortest path based on the link weight of Fig.3(d). Because the flows pass through the SDN node can flexibly choose the outgoing links regardless of their weights, we use dotted lines to represent the extra possible links that the SDN node can split the flows to. In Fig.3(f), we assume that a new node D joins in the network and new traffic flow $T4$=(D,B,2) is generated. Thus the injected flow on SDN node A to destination B is 2 units. The algorithm calls \emph{Flow Allocation} Function to calculate traffic splitting ratio for Node A. Since the load utilization of path (A,B) is zero, and load utilization of path (A-C-B) is 0.4, the function prefers to choose the high-utilization path (A-C-B) carrying $T4$. At last, the traffic load of link (A,B), (A,C) and (C,B) is 0 unit, 6 units and 6 units, respectively. The idle link (A,B) can be turned off for saving energy in the network.
\begin{figure}[t]
\begin{center}
\includegraphics[width=8.8cm]{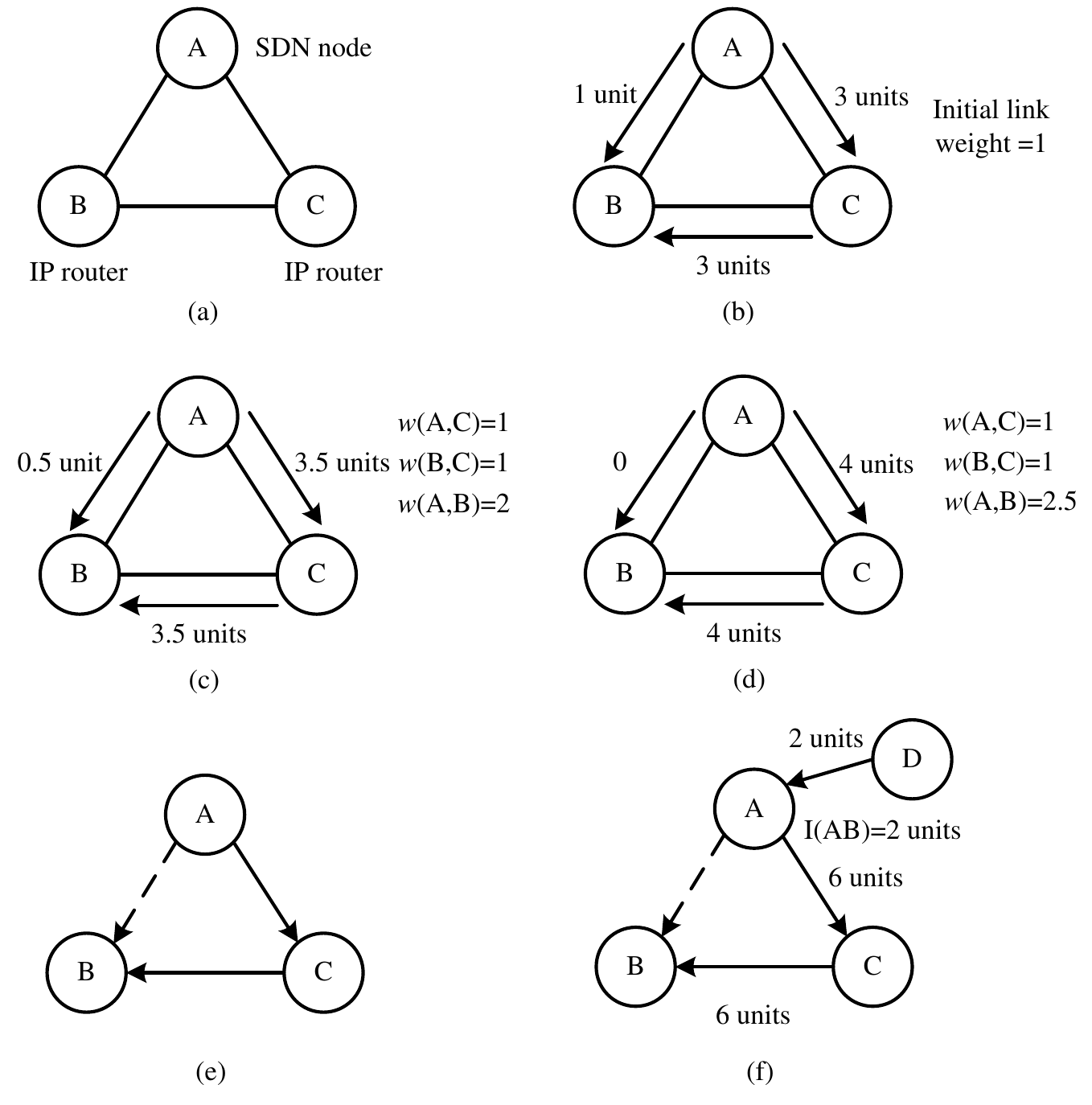}
\caption{A simple example of the algorithm (a) network topology (b) link load based on initial link weight (c) link load based on adjusted link weight (Iteration 1) (d) link load based on adjusted link weight (Iteration 2) (e) the shortest path from other nodes to node B (f) traffic splitting ratio calculation of SDN node} 
\label{Figure:HEATE_example}
\end{center}
\end{figure}

\vspace{10pt}
\section{Simulation Results}
\label{Section:simulation}
Simulation in this section is implemented in NS2, using C++ and OTcl as the programming language. The simulation is carried out on two network topologies including the Geant network which has 23 nodes and 74 links and the Sprintlink topology which has 30 nodes and 138 links from Rocketfuel Project [28]. For the two network topologies, we use the following method to set link capacity. All of the network nodes are divided into POP1 nodes and POP2 nodes. The POP1 node's degree is less than 3; and the POP2 node's degree is not less than 3. If the link connecting one POP1 node, we set the link capacity as 2.5Gbps; if the link connecting two POP2 nodes, we set the link capacity as 10Gbps.
\par 
In our work, to approve algorithm effectiveness we compare our proposed HEATE algorithm with pure Energy-Aware OSPF (EA-OSPF) algorithm and Energy-Aware Flow Allocation (EA-FA) algorithm. To improve energy efficiency in the hybrid network, EA-OSPF only performs optimizing OSPF link weight, and EA-FA algorithm only performs optimizing traffic splitting ratio in SDN-enabled switches. For all these three algorithms, the traffic matrix is generated as Eq.(12) [29], where $d_{ij}$ denotes the traffic flow from node $i$ to node $j$, $C_{ij}$ denotes the link capacity of link $(i,j)$, and $\sigma_i$ denotes the random figure between 0 and 0.1. We randomly generate 50 traffic matrices.
\begin{equation}
d_{ij}=\sigma_i\sum_{\{t|(i,t)\in L\}} C_{it} \frac{\sum_{\{t|(t,j)\in L\}} C_{tj}}{\sum_{\{(m,n)|(m,n)\in L\}} C_{mn} - \sum_{\{t|(i,t)\in L\}} C_{it}}.
\end{equation}

Fig.4 and Fig.5 show energy saving ratio vs number of SDNs for Geant network and Sprintlink network, respectively. The definition of energy saving ratio in our study is the ratio of number of turned-off links to total number of links. From Fig.4 and Fig.5, we can see that energy saving ratio of HEATE and EA-FA increases rapidly with the increase of deployed SDN-enabled switches in the network. The reason is that with the increase of SDNs, the central SDN controller can control more traffic flows and globally choose the optimal path for them to maximize energy saving. When the deployment of SDNs is greater than a threshold (i.e., 16 nodes in the Geant network and 21 nodes in the Sprintlink network), the variation of energy saving ratio becomes relatively flat. This is because when the deployment of SDNs reaches a threshold, an approximately optimal energy-saving flow allocation can be achieved. More SDN-enabled switches deployed into the network will make no obvious enhancement in energy saving. We demonstrate that when deploying the SDN at a small rate, we can obtain the most of the energy efficient benefit. Since EA-OSPF achieves energy efficiency by link weight optimization, the increase of number of SDNs does not influence energy saving ratio of EA-OSPF.

\begin{figure}[t]
\begin{center}
\includegraphics[width=8.7cm]{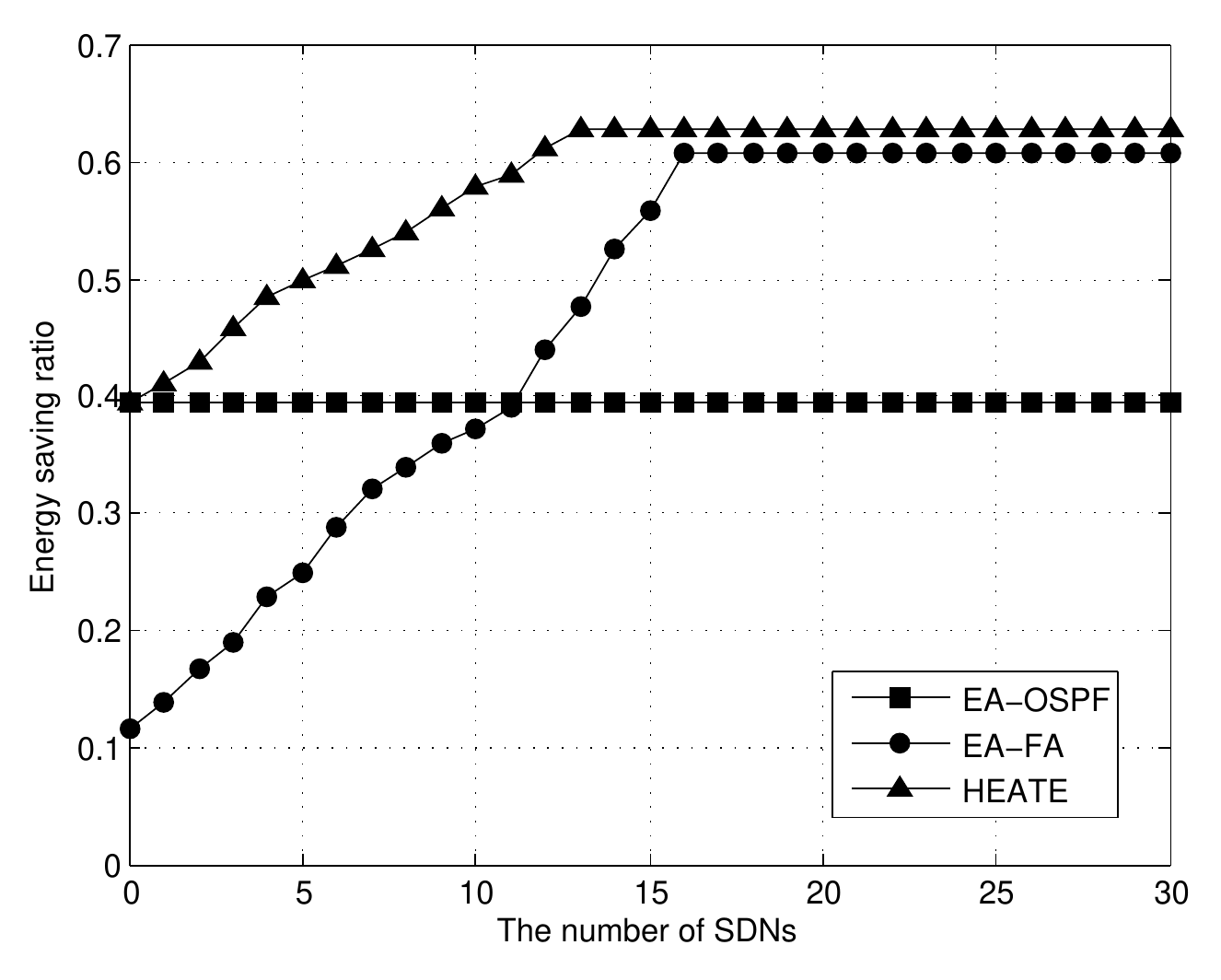}
\caption{Energy saving ratio vs number of SDNs for Geant network} 
\label{Figure:Sim_es_no_geant}
\end{center}
\end{figure}

\begin{figure}[t]
\begin{center}
\includegraphics[width=8.7cm]{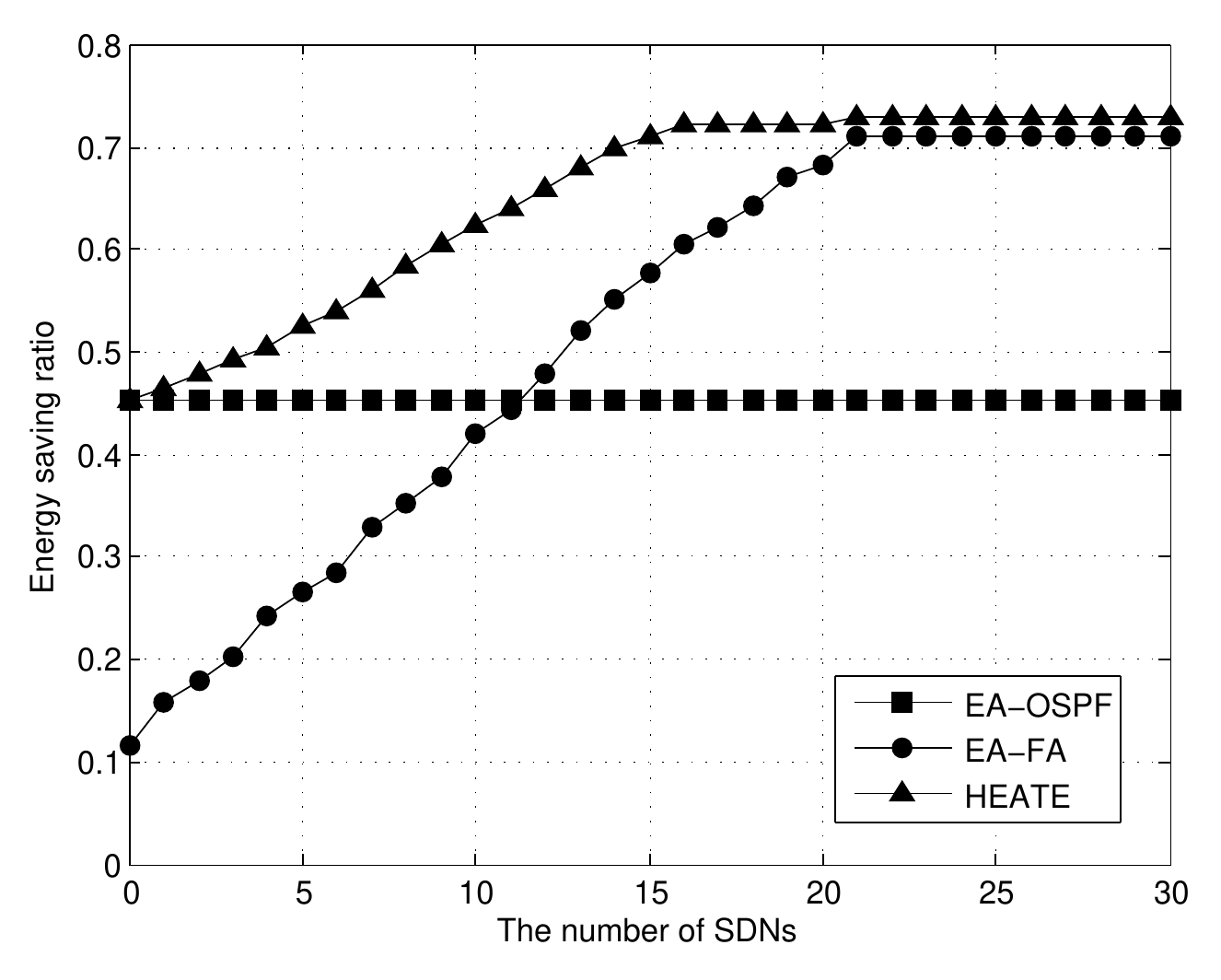}
\caption{Energy saving ratio vs number of SDNs for Sprintlink network} 
\label{Figure:Sim_es_no_sprint}
\end{center}
\end{figure}

Fig.6 and Fig.7 show energy saving ratio of three algorithms under different traffic matrices for Geant network and Sprint link network, respectively. We assume that the number of SDN-enabled switches is 6 in the hybrid network.  From Fig.6 and Fig.7, we can see that HEATE saves more 13.2\% energy consumption compared with EA-OSPF, and more 21.7\% energy consumption compared with EA-FA in average. This is because that the HEATE jointly optimizes the OSPF link weight and traffic flow splitting ratio in the hybrid network. The above two procedures are interactive within the algorithm. Thus the HEATE aggregates the uncontrollable flow and controllable flow onto partial links and turns off idle links. EA-OSPF only optimizes OSPF link weight, and EA-FA only optimizes traffic splitting ratio at SDNs. So the energy-saving effect of the latter two algorithms is not as good as that of the HEATE.

\begin{figure}[t]
\begin{center}
\includegraphics[width=8.7cm]{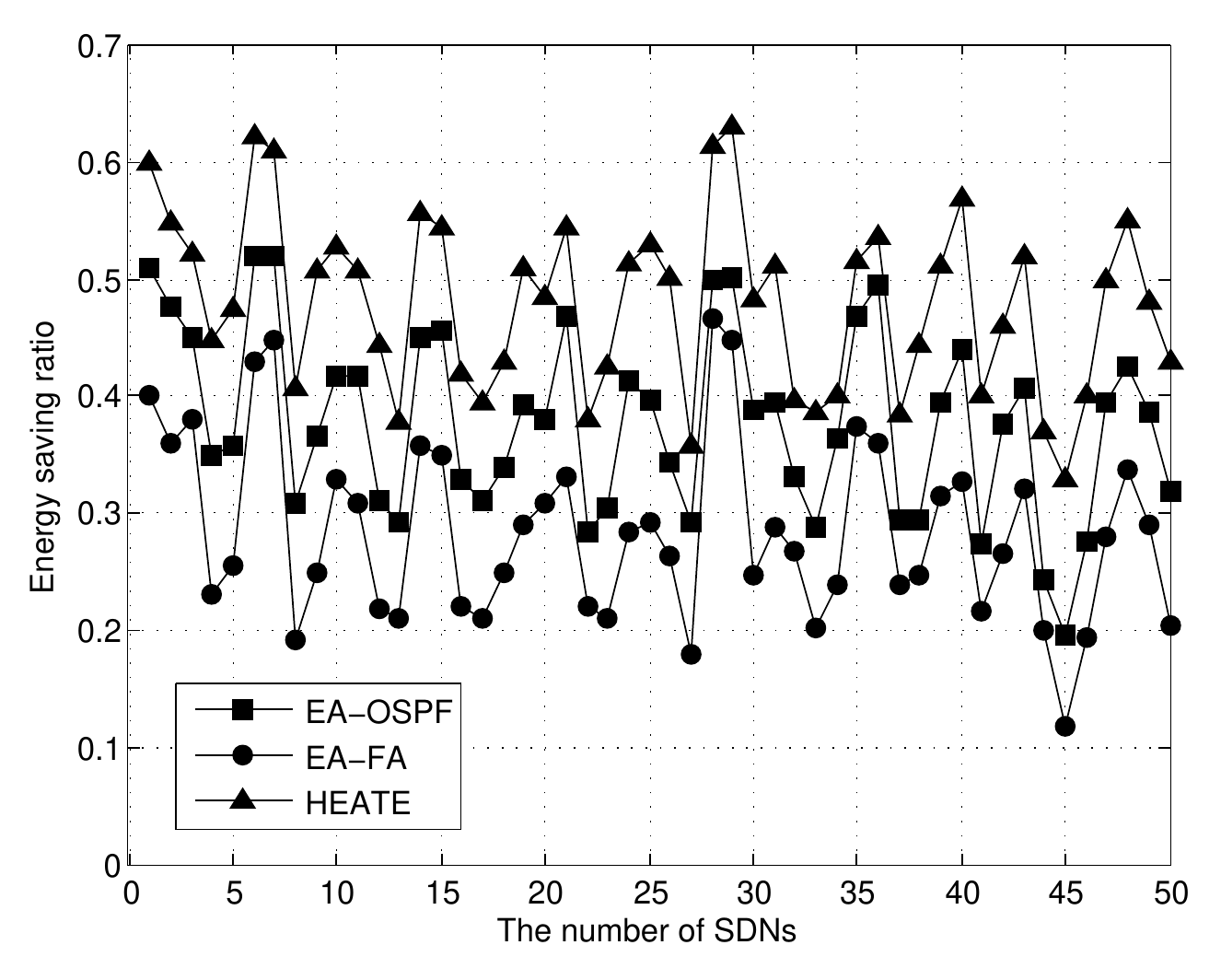}
\caption{Energy saving ratio under different traffic matrices for Geant network} 
\label{Figure:Sim_es_traffic_matrices_geant}
\end{center}
\end{figure}

\begin{figure}[t]
\begin{center}
\includegraphics[width=8.7cm]{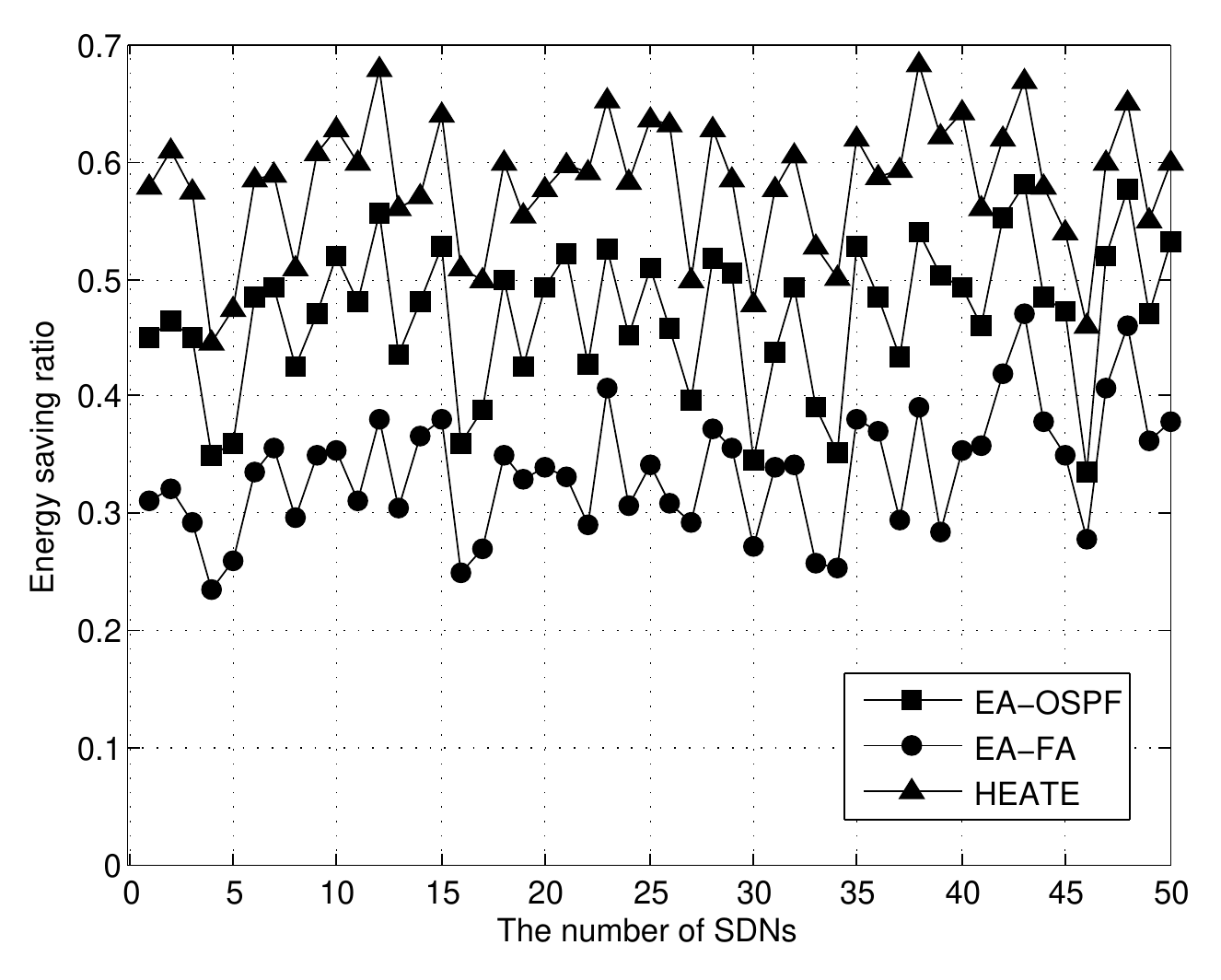}
\caption{Energy$\!$ saving$\!$ ratio under different traffic$\!$ matrices for Sprintlink$\!$ network} 
\label{Figure:Sim_es_traffic_matrices_sprint}
\end{center}
\end{figure}

\vspace{10pt}
\section{\uppercase{Conclusion}}
\label{Section:conclusion}
In this paper, we study on energy-efficient traffic engineering problem in hybrid SDN/IP networks. Since the problem to solve in our work is NP-hard, we propose a novel heuristic HEATE (Hybrid Energy-Aware Traffic Engineering) algorithm to solve this problem. The HEATE algorithm jointly optimizes OSPF link weight of IP routers and traffic flow splitting ratio of SDN-enabled switch. Thus traffic flow is aggregated onto partial links and the underutilized links can be turned off to save energy. By computer simulation results, we show that our algorithm has a significant improvement in energy efficiency in hybrid SDN/IP networks.

\bibliographystyle{jcn}

\begin{thebibliography}{10}
\bibitem{McAn08}
N. McKeown, T. Anderson, H. Balakrishnan, G. Parulkar, L. Peterson, J. Rexford, S. Shenker, J. Turner, "OpenFlow: Enabling Innovation in Campus Networks", ACM SIGCOMM CCR, 2008.

\bibitem{HoKa13}
C.-Y. Hong, S. Kandula, R. Mahajan, M. Zhang, V. Gill, M. Nanduri, and
R. Wattenhofer, Achieving high utilization with software-driven wan",
in Proceedings of the ACM SIGCOMM 2013 conference on SIGCOMM.
ACM, 2013, pp. 15-26.

\bibitem{NS09}
Handigol N, Seetharaman S, Flajslik M, et al, "Plug-n-Serve: Load-balancing web traffic using
OpenFlow", ACM SIGCOMM Demo, 2009.

\bibitem{JaKu13}
S. Jain, A. Kumar, S. Mandal, J. Ong, L. Poutievski, A. Singh,
S. Venkata, J. Wanderer, J. Zhou, M. Zhuet al., B4: Experience with
a globally-deployed software defined wan", inProceedings of the ACM
SIGCOMM 2013 conference on SIGCOMM. ACM, 2013, pp. 3-14.

\bibitem{BJ02}
Fortz  B,  Rexford  J,  Thorup  M,  "Traffic  engineering  with  traditional  IP  routing  protocols",
IEEE Communications Magazine, vol. 40, no. 10, pp. 118-124, 2002.

\bibitem{INFOCOM2010_es_ospf}
A. Cianfrani, V. Eramo, M. Listanti, M. Marazza, and E. Vittorini, "An energy saving routing algorithm for a green OSPF protocol", in Proc. IEEE INFOCOM, Mar. 15-19, 2010, pp. 1-5.

\bibitem{ICM2010_review_mobile_radio_networks}
L. M. Correia, D. Zeller, O.Blume, D. Ferling, Y. Jading, I. Godor,G. Auer, and L. Van der Perre, "Challenges and enabling technologies for energy aware mobile radio networks", IEEE Commun. Mag., vol. 48, no. 11, pp. 66-72, Nov. 2010.

\bibitem{GreenComm2009_es_cellular}
M. A. Marsan, L. Chiaraviglio, D. Ciullo, and M. Meo, "Optimal energy savings in cellular access networks", in Proc. Green-Comm09, Jun. 14-18, 2009, pp. 1-5.

\bibitem{ICM2011_green_backbone_networks}
W. Fisher, M. Suchara, and J. Rexford, "Greening backbone networks: Reducing energy consumption by shutting off cables in bundled links", in Proc. 1st ACM SIGCOMM Workshop Green Netw., Aug. 2010, pp. 29-34.

\bibitem{ICM2011_es_wired_wireless_networks}
J. Baliga, R. Ayre, K. Hinton, and R. Tucker, "Energy consumption in wired and wireless access networks", Communications Magazine, IEEE , vol.49, no.6, June 2011.

\bibitem{IN2011_backbone_sleep}
R. Bolla, R. Bruschi, A. Cianfrani, and M. Listanti, "Enabling backbone networks to sleep", Network, IEEE, vol.25, no.2, 2011.

\bibitem{ICST_es_future_internet}
R. Bolla, R. Bruschi, F. Davoli, and F. Cucchietti, "Energy Efficiency in the Future Internet: A Survey of Existing Approaches and Trends in Energy-Aware Fixed Network Infrastructures", Communications Surveys and Tutorials, IEEE, vol.13, no.2.

\bibitem{CC2015_sdn_es}
H. Peng, W. Wang, C. Wang, etc., "A SDN-based energy saving strategy in wireless access networks", China Communications, Vol. 12, no. 8, 2015, pp.132-145.

\bibitem{Globecom2014_rule_sdn}
F. Giroire, J. Moulierac, T. K. Phan, "Optimizing rule placement in software-defined networks for energy-aware routing", IEEE Globecom 2014, pp. 2523-2529.

\bibitem{IOP_2008}
J. G. Koomey, "Worldwide electricity used in Data Centres", IOP Publishing Ltd, 2008.

\bibitem{IBMJRD2009_es_data_center}
H. Hamann, T. van Kessel, M. Iyengar, J.-Y. Chung, W. Hirt, M. Schappert, A. Claassen, J. Cook, W. Min, Y. Amemiya, V. Lopez, J. Lacey, and M. O'Boyle, "Uncovering energy-efficiency opportunities in data centers", IBM Journal of Research and Development. vol.53, no.3, 2009.

\bibitem{IT2011_green_internet}
A.P. Bianzino, A.K. Raju, D. Rossi, "Greening the Internet: Measuring Web Power Consumption", IT Professional, vol.13, no.1, pp.48-53, Jan.- Feb. 2011.

\bibitem{IBMCARE2012_es_datacenter_sdn}
D. Kakadia and V. Varma, "Energy efficient data center networks-A SDN based approach", IBM Collaborative Academia Research Exchange, 2012.

\bibitem{EWSDN2012_es_openflow}
L. Prete, F. Farina, M. Campanella, and A. Biancini, "Energy efficient minimum spanning tree in OpenFlow networks", In Proc. Eur. Workshop Softw. Defined Netw., Oct. 25-26, 2012, pp. 36-41.

\bibitem{AnKo13}
S. Agarwal, M. Kodialam, and T. Lakshman, Traffic engineering in software defined networks", in IEEE INFOCOM, 2013.

\bibitem{LeCa13}
D. Levin, M. Canini, S. Schmid, and A. Feldmann, Panopticon: Reaping the Benefits of Partial SDN Deployment in Enterprise
Networks," Technical report, U Berlin / T-Labs, 2013.

\bibitem{ViVa14}
S. Vissicchio, L. Vanbever, L. Cittadin, and et al. Safe Updates of Hybrid SDN Networks", ACM-SIGCOMM HotSDN
Workshop, 2014.

\bibitem{RE11}
Marcelo R. Nascimento, Christian E. Rothenberg, etc, "Virtual routers as a service: the routeflow approach leveraging software-defined networks", Proceedings of the 6th International Conference on Future Internet Technologies, 2011.

\bibitem{ChSo08}
J. Chabarek, J. Sommers, P. Barford, C. Estan, D. Tsiang, and
S. Wright, "Power awareness in network design and routing",
in INFOCOM, 2008.

\bibitem{GoHu60}
R.  Gomory,  T.  Hu,  "Multi-terminal  network  flows",  Journal  of  the
Society for Industrial  Applied Mathematics, vol. 9, no. 4, pp. 551-570,
1960.

\bibitem{BM10}
Fortz B, Thorup M, "Internet traffic engineering by optimizing OSPF weights", In IEEE Infocom, 2010. 

\bibitem{Hoff97}
Michael E. Hoffman,"The Algebra of Multiple Harmonic Series", Journal of Algebra, vol. 194, no. 2, pp. 477-495, 1997.

\bibitem{TeMa03}
Renata Teixeira, Keith Marzullo, Stefan Savage, "Characterizing and measuring path diversity of internet topologies", Proceedings of the ACM SIGMETRICS, 2003.

\bibitem{Globecom2011_ERMAO}
Li K, Wang S, Xu S, et al., "ERMAO: An Enhanced Intradomain Traffic Engineering Approach in LISP-Capable Networks", Globecom 2011, 2011 IEEE. 2011:1-5.

\end{thebibliography}

\end{document}